\documentclass[10pt]{article}

\usepackage{amsmath,amsfonts,amssymb}
\usepackage{graphicx}
\usepackage{subfigure}
\usepackage{dcolumn}
\usepackage{bm}
\usepackage{longtable}

\newcommand{\Vd}{\boldsymbol V}
\newcommand{\nd}{\boldsymbol n}

\usepackage{graphicx}

\usepackage{cite}

\usepackage{color}

\usepackage[labelfont=bf,labelsep=period,justification=raggedright]{caption}

\makeatletter
\renewcommand{\@biblabel}[1]{\quad#1.}
\makeatother

\date{}

\newcommand{\diff}[2]{\frac{\mathrm{d}#1}{\mathrm{d}#2}}

\begin{document}

\begin{flushleft}

{\Large \textbf{A globally attractive cycle driven by sequential bifurcations containing ghost effects in a 3-node yeast cell cycle model} }
\\
Fangting Li$^{1,2,\ast}$,
Mingyang Hu$^{2,3}$,
Bo Zhao$^{1}$,
Hao Yan$^{1}$,
Bin Wu$^{1}$,
Qi Ouyang$^{1,2,\ast}$
\\
\bf{1} School of Physics, Peking University, Beijing 100871, China
\\
\bf{2} Center for Quantitative Biology, Peking University, Beijing 100871, China
\\
\bf{3} Yuanpei College, Peking University, Beijing 100871, China
\\
$\ast$ E-mail: li\_fangting@pku.edu.cn, qi@pku.edu.cn
\end{flushleft}


\section*{Abstract}

Yeast cells produce daughter cells through a DNA replication and mitosis cycle associated with checkpoints and governed by the cell cycle regulatory network. To ensure genome stability and genetic information inheritance, this regulatory network must be dynamically robust against various fluctuations. Here we construct a simplified cell cycle model for a budding yeast to investigate the underlying mechanism that ensures robustness in this process containing sequential tasks (DNA replication and mitosis). We first establish a three-variable model and select a parameter set that qualitatively describes the yeast cell cycle process. Then, through nonlinear dynamic analysis, we demonstrate that the yeast cell cycle process is an excitable system driven by a sequence of saddle-node bifurcations with ghost effects. We further show that the yeast cell cycle trajectory is globally attractive with modularity in both state and parameter space, while the convergent manifold provides a suitable control state for cell cycle checkpoints. These results not only highlight a regulatory mechanism for executing successive cell cycle processes, but also provide a possible strategy for the synthetic network design of sequential-task processes.

\section*{Author Summary}
A challenge in biology is to understand how complex regulatory networks in a cell execute different biological functions, such as the DNA replication event and mitosis event in eukaryotic cell cycle processes, orderly and reliably. We call the process as the sequential-task process. In this work, we establish a simplified toy model for a budding yeast cell cycle, and develop the analysis methods to describe the evolving process. Our results show that the cell cycle process is an excitable system with global robustness towards different initial states, where the phase transitions of the cell cycle are controlled by feedback in the network and driven by bifurcations with a slow passage that we refer to as a ghost effect. These ghost effect bifurcations are sufficient for providing both the longer duration of each event and the modularity in state/parameter space. The modularity and decoupling of the kinetic parameters in different events ensures the stability of the system against parameter changes; thus small parameter changes in any given event should not influence the general character of the cell cycle process. We describe this as the wave transition ``domino'' model of the cell cycle. This wave transition model not only provides an efficient and easily controlled strategy for the cell cycle, but also suggests a guide for building synthetic networks of sequential-task processes.

\section*{Introduction}

It has been suggested that dynamic robustness and modularity are the important characteristics of a cellular regulatory network for fulfilling complex biological processes. Recent progress in quantitative modeling and experiments of some fundamental cellular processes, including the
chemotaxis in E. coli \cite{Leibler1997, Wingreen2008} and the cell cycle process in yeast and oocyte \cite{Chen2000, Li2004, Chen2004, Cross2010, Murray&Kirschner1989,Ferrell2009}, have provided a better systematic and quantitative view of these biological processes. These studies have revealed the structural and dynamic properties of regulatory networks, such as the basic motifs and modules of network structure \cite{Alon2002, Alon-2007, Alon2007}, the feedback loops that dynamically control the genetic switches and transitions\cite{Collins2003, Ferrell2003, Alexander2004, Cross2008}, and the dynamic stability and robustness of such networks\cite{Leibler1997, Ma2009}.

In the cell cycle process, the cell successively executes DNA replication and mitosis events using a checkpoint mechanism. The dynamic regulatory mechanism in cell cycle processes was revealed by the pioneering work of Tyson, Novak, and Ferrell \cite{Tyson2001, Tyson-BJ-2006, Ferrell2011}. The cell cycle process is considered to be a series of irreversible transitions from one state to another \cite{Morgan, Novak-NCB-2007}, which is regulated by system-level feedback \cite{Novak-NCB-2007}. For example, the cell cycle process in oocyte maturation is built upon a hysteretic switch with two saddle-node bifurcations \cite{Ferrell2009}. The skeleton models of the cell cycle in a budding yeast \cite{Tyson2013} and mammalian cells \cite{Goldbeter2011} have also been established to investigate the temporal order and sequential activation in the cell cycle. Furthermore, a DNA replication checkpoint in a mammalian cell cycle has been proposed to lengthen the cell cycle period and lead to a better separation of the S and M phases \cite{Goldbeter2009}.

To reveal the essential dynamic properties of a sequential-task cell cycle process, some fundamental questions must be studied. For instance, how can the cell cycle network ensure the cell reliably executes a successive order of events?  How does the network ensure that the process is robust against environmental fluctuations? What is the function of the cell cycle checkpoint from a dynamic view? Here, we address these questions using a simplified cell cycle network model in a budding yeast \textsl{Saccharomyces cerevisiae}. We first develop a simplified yeast cell cycle model and select a set of parameters that can qualitatively describe the cell cycle process. We then apply perturbation methods to analyze the dynamic properties of the local manifold along the evolving cell cycle trajectory. We demonstrate the following: that the yeast cell cycle process is an excitable system driven by sequential saddle-node bifurcations; the yeast cell cycle trajectory is globally attractive with modularity in both state space and parameter space; and that the convergent manifold of the trajectory corresponds to the cell cycle checkpoints. With these advantages, the yeast cell cycle process becomes effective and reliable throughout its execution.
\section*{Model}

The budding yeast cell cycle process is a well-established system for investigating and exploring the control mechanisms of regulatory networks \cite{Chen2000, Chen2004,Li2004, Cross&Siggia2005, Tyson2006}. Recent works integrating modeling and experiment have highlighted the function of positive and negative feedback in the Start point \cite{Cross2008, Cross2010, Yang2013}, spindle separation process \cite{Morgan2008}, and mitotic exit \cite{Novak&Uhlmann2009, Lu&Cross2010}. To investigate the essential dynamic features of budding yeast cell cycle, especially the dynamic robustness and modularity, we construct and analyze a coarse-graining model that is mainly based on the abstract architectures and functions of cell cycle and ignored the molecular details. These simplification approaches have been applied widely to study the design principles of regulatory networks \cite{Ma2006, Ma2009, Lim2013}

\subsection*{The simplified yeast cell cycle network and model}

The cell cycle process in a budding yeast produces new daughter cells through two major events: DNA replication in the S phase and mitosis in the M phase. We denote this kind of processes that execute different biological functions and events in order and reliably as the sequential-task processes. The different cyclins and transcription factors (TF) control different successive events, and the DNA replication checkpoint and spindle assembly and separation checkpoint ensure the stability and hereditary validity of the genetic information \cite{Morgan}. The checkpoints ensure the completion of earlier events before the beginning of later events to maintain the orderly progression of the cell cycle \cite{Hartwell}.

\begin{figure}[!ht]
\begin{center}
\scalebox{0.2}{\includegraphics{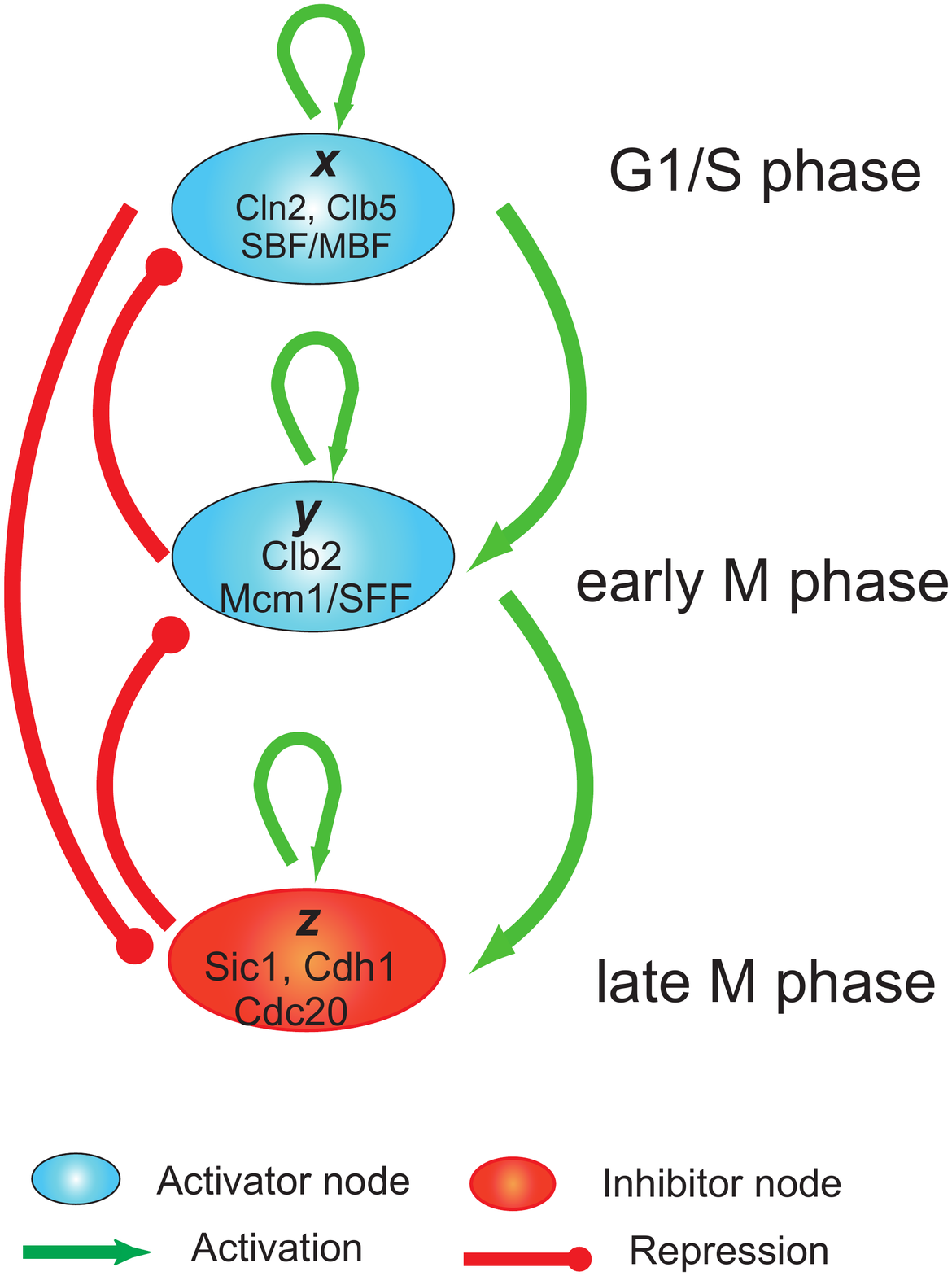}}
\caption{\label{network} The simplified regulatory network in a budding yeast cell-cycle, where
$x$, $y$, and $z$ represent key regulators of the G1/S, early M, and late M modules, respectively. Different modules are connected by activation and repression interactions. For details, see the main text and Supporting Information. }
\end{center}
\end{figure}

Based on the cell cycle regulatory network (Figure 3-34 in \cite{Morgan}), our previous network with an intra-S and spindle checkpoint \cite{Li2004}, and the full cell cycle network in the supplemental information (SI), we separated the yeast cell cycle regulatory network into the G1/S, early M and late M modules, where the positive and negative feedback loops play an essential role in governing the yeast cell cycle process. We then obtained a coarse-graining simplified 3-module network, shown in Figure~\ref{network}. In the figure, $x$ represents the activities of key regulators such as the cyclins Cln2 and Clb5, and SBF and MBF in the G1 and S phases; $y$ represents the activities of key regulators such as the cyclin Clb2 and the transcriptional factor Mcm1/SFF in the early M phase; and $z$ represent the activities of key inhibitors such as Cdh1, Cdc20, and Sic1 in the late M/G1 phase. For simplification, we ignore the G2 phase.

Then a simplified three-variable ordinary differential equation (ODE) model can be established. The 3-node cell cycle model is not only reduced on the topological structure of budding yeast cell cycle network, but also constructed to describe the key dynamic features of yeast cell cycle, especially the genetic switches, irreversible transitions, and the sequential-task process from DNA replication to mitosis. The 3-node model is written as follows:
\begin{subequations}
\begin{align}
\diff{x}{t}=&\frac{x^2}{j_{1}^2+x^2}-k_1x-yx, \\
\diff{y}{t}=&k_{a1}x+\frac{y^2}{j_{2}^2+y^2}-k_2y-zy, \\
\diff{z}{t}=&k_{a2}y+\frac{k_sz^2}{j_{3}^2+z^2}-k_3z-k_ixz.
\end{align}
\label{nondimensionalequ}
\end{subequations}

In this 3-node model, we assume that the simplest forms to represent the interactions in the cell cycle network in order to qualitatively describe the genetic switches in the START point, G2/M transition and the late M phase, and we introduce the checkpoint mechanism by changing kinetic parameters. The positive feedback in each module is represented by a 2-order Hill function, while the inhibition and repression component among different modules is assumed to have its simplest form, i.e., $-yx$, $-zy$, and $-xz$. The parameters $j_{i}$ and $k_{i}$ are the Hill coefficients and degradation rates of $x$, $y$ , and $z$ respectively; $k_{a1}x$ is the activation rate from $x$ to $y$; and $k_{a2}y$ is the activation rate from $y$ to $z$. Furthermore, the $-k_ixz$ term in Eq. \ref{nondimensionalequ}c represents the repression from $x$ to $z$, $-zy$ in Eq. \ref{nondimensionalequ}b represents the repression from $z$ to $y$, and $-yx$ in Eq. \ref{nondimensionalequ}a represents the repression from $y$ to $x$, these are negative feedback. More detail concerning network reduction, model assumptions, formulations, and the dimensionless form deduction can be found in SI.

We focus on the dynamic robustness and modularity of the cell cycle process, particularly from the excited G1 state to the final steady G1 state. Before we quantitatively analyze the model, we qualitatively illustrate  the dynamics of $x$, $y$, and $z$ and the corresponding yeast cell cycle events and process. At the beginning of the cell cycle process in the G1 phase state, inhibitor $z$ is at a high level with low levels of $x$ and $y$, and the cell is in the resting G1 state. Once the system enters the cell cycle process, $x$ is activated and dominant, which represses the inhibitor $z$ to a lower level. In the early M phase, $y$ is activated and dominant, and the activated $y$ represses $x$ in turn. In the late M and G1 phases, a $z$ wave is triggered by $y$ and increases to high levels. Sequentially, the x wave is dominant in the S phase, the y wave is dominant in the early M phase, and the z wave is dominant in the late M and G1 phases. We denote this set of sequential steps as the wave transition ``domino" model of the yeast cell cycle process.

The excited G1 state of the yeast cell cycle process can be settled as follows. If we set $k_{a1}=k_{a2}=0$ and ignore the repression terms $-xy$, $-yz$, and $-k_ixz$ in Eq. \ref{nondimensionalequ}, the equations for $x$, $y$, and $z$ can be decoupled. In this situation, the steady state of $x$ in equation $\frac{dx}{dt}=x^2/(j_{1}^2+x^2)-k_1x \equiv G(x,j_1,k_1)=0$ has three fixed points, i.e., $ x^{(1)}=0,\
x^{(2)}=\frac{1-\sqrt{1-4j^{2}_{1}k^{2}_{1}}}{2k_1}$, and
$x^{(3)}=\frac{1+\sqrt{1-4j^{2}_{1}k^{2}_{1}}}{2k_1} $, where $x^{(1)}$ and $x^{(3)}$ are stable nodes and $x^{(2)}$ is an unstable saddle that serves as the threshold to the excited G1 state and separates the attractive ranges of $x^{(1)}$ and $x^{(3)}$. These dynamic features are only determined by $j_1$ and $k_1$. Similarly,
if we set $H(y,j_2,k_2)\equiv \frac{y^2}{j_{2}^2+y^2}-k_2y=0$ and $I(z,k_s,j_3,k_3)\equiv \frac{k_sz^2}{j_{3}^2+z^2}-k_3z=0$, we have $ y^{(1)}=0, \ y^{(2)}=\frac{1-\sqrt{1-4j^{2}_{2}k^{2}_{2}}}{2k_2},
\ y^{(3)}=\frac{1+\sqrt{1-4j^{2}_{2}k^{2}_{2}}}{2k_2}, $ $
z^{(1)}=0,\
z^{(2)}=\frac{k_s-\sqrt{k_s^2-4j^{2}_{3}k^{2}_{3}}}{2k_3}$, and
$z^{(3)}=\frac{k_s+\sqrt{k_s^2-4j^{2}_{3}k^{2}_{3}}}{2k_3}.
$ The excited G1 state is defined as $(x_0 \geq x^{(2)}, y_0= y^{(1)}, z_0=z^{(3)})$.

\subsection*{Constraint conditions and suitable kinetic parameters}

In the yeast cell cycle process, the cell should finish the DNA replication before mitosis, and there should be enough time to execute both the DNA replication and mitosis events. The question then arises as to what criteria are available to qualitatively describe the yeast cell cycle process. We use an event order and the long duration of the $x$ and $y$ waves as constraints to search for suitable kinetic parameter sets in Eq. \ref{nondimensionalequ}. The first constraint is that the $x$ wave is followed by the $y$ wave, while the $z$ wave decreases in the S phase and increases in the late M and G1 phases. The second constraint is that the duration times (calculated as the full duration at half maximum) of both the $x$ and $y$ waves should be sufficiently long (i.e., $\geq 30$ arbitrary units (au)) in order to ensure the successful execution of any DNA replication and mitosis events.

\begin{figure}[!ht]
\begin{center}
\centerline{\includegraphics[width=0.8\textwidth]{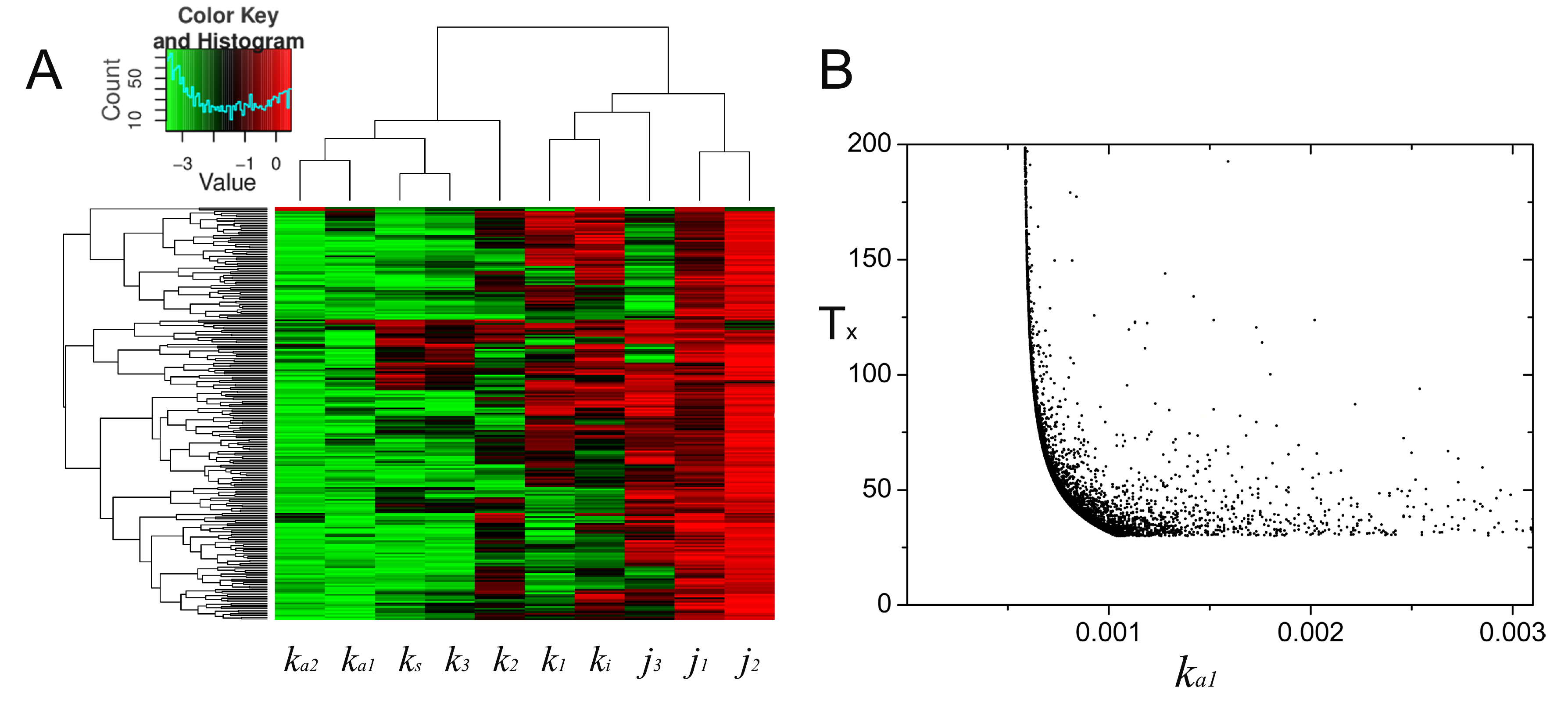}}
\caption{\label{clustering01}The durations of the G1/S and early M phases are controlled the activation rates $k_{a1}$ and $k_{a2}$, respectively. (A) Clustering map of the parameter sets in log scale that satisfy the fundamental constraint conditions of the yeast cell-cycle process. The insert gives the scale of value. (B) The $x$ wave duration $T_x$ of the satisfied parameter sets as a function of $k_{a1}$. The envelope curve corresponds to $T_x\propto (k_{a1}-k^c_{a1})^{-1/2}$.}
\end{center}
\end{figure}

Using the $10^{7}$ random parameter sets we produced by Latin hypercube sampling \cite{Numbericalrecipes} and the excited G1 state as initial state, we obtained 239 groups of parameter sets in a simulation iteration that satisfied the above conditions. These parameter sets were clustered as shown in Figure \ref{clustering01}A. We found that the above constraint conditions only required small activation rate values for $k_{a1}$ (from the $x$ wave to the $y$ wave) and $k_{a2}$ (from the $y$ wave to the $z$ wave). Furthermore, $k_s$ and $k_3$---which are related to $z$---must be correlated, while $j_1$, $j_2$, and $j_3$ are relatively large in value. Different simulation iterations showed similar results.

We then set $j_m=0.5$ and $k_m=0.2$ ($m=1,2,3$) to search for the satisfied parameters $k_{a1}$, $k_{a2}$, $k_i$, and $k_s$. The results are plotted in Figure \ref{clustering01}B, with the $x$ wave duration $T_x$ as a function of $k_{a1}$. The envelope curve corresponds to $T_x\propto (k_{a1}-k^c_{a1})^{-1/2}$, where $k^c_{a1}$ will be discussed in the following section. Furthermore, a suitable value of $k_{a1}$ is $0.001$ for the constraint $T_x > 30\ au$. For the same reason, we select $k_{a2} = 0.001$. These results give us the parameter set that can qualitatively describe the yeast cell cycle trajectory: $j_m=0.5$, $k_m=0.2$ ($m=1,2,3$), $k_{a2} = k_{a2} = 0.001$, $k_i=5.0$, and $k_s=1.0$. We denote this set as the perfect yeast cell cycle parameter set.

\subsection*{Duration of G1/S and early M phases is controlled by strength of activation through the ghost effect}

What is the relationship between the duration of the $x$ or $y$ wave and $k_{a1}$ or $k_{a2}$? To illustrate this relationship, we analyze how the $x$ wave triggers the activation of the $y$ wave. During the late S phase and early M phase, $x$ is almost fully activated and has repressed $z$ to zero, so we can ignore $z$ in Eq. \ref{nondimensionalequ} to get

\begin{subequations}
\begin{align}
\cfrac{dx}{dt}=&\frac{x^2}{j_{1}^2+x^2}-k_1x-xy=G(x,j_1,k_1)-xy, \\
\cfrac{dy}{dt}=&k_{a1}x+\frac{y^2}{j_{2}^2+y^2}-k_2y=k_{a1}x+H(y,j_2,k_2).
\end{align}
\label{xyequations}
\end{subequations}
We showed in the previous section that $G(x,j_1,k_1)$ and $H(y,j_2,k_2)$ determine $x^{(i)}$ and $y^{(i)}$ ($i=1,2,3$), so the activation term $k_{a1}x$ in Eq. \ref{xyequations}b will trigger the activation of $y$, and the repression term $-xy$ in Eq.~\ref{xyequations}a will repress $x$. In the early M phase, $k_{a1}x$ triggers the activation of $y$ through a saddle-node bifurcation, where two smaller fixed points of $y$ ($y^{(1)}$ and
$y^{(2)}$) collide and vanish, leaving only $y^{(3)}$. We denote $k_{a1}^c$ as the critical bifurcation point of $k_{a1}$, and its value can be estimated as follows: $H(y,j_2,k_2)\equiv\frac{y^2}{j_{2}^2+y^2}-k_2y$ is defined as part of the right hand side terms of Eq.~\ref{xyequations}b, while $H(y^*)$ is the minimum of $H(y,j_2,k_2)$. We can ignore the inhibition term $-yx$ in Eq.~\ref{xyequations}a in the late S phase because $x$ is approaching its maximum $x^{(3)}$ and $y$ is not fully activated. When $k_{a1}x^{(3)}\geq |H(y^*)|$ or
$k_{a1}\geq k_{a1}^c\equiv |H(y^*)|/x^{(3)}$, the saddle-node bifurcation will occur. Note that $x^{(3)}$ is only determined by $j_1$ and $k_1$, while $H(y^*)$ is determined by $j_2$ and $k_2$. Thus, $k_{a1}^c$ is determined by $j_1$, $k_1$, $j_2$, and $k_2$.

Furthermore, if $0<(k_{a1}-k_{a1}^c)\ll 1$, then just after the bifurcation we have $x \simeq x^{(3)}$, $y \simeq y^*\simeq y^{(1)}$, and $z
\simeq 0$, so
$$
\diff{y}{t}= k_{a1}x+H(y,j_2,k_2)\simeq (k_{a1}-k^c_{a1})x^{(3)}+c(y-y^{(1)})^2 \ll 1.
$$
It is just after the two smaller fixed points collide that the saddle-node remnant, or ghost effect, with a slow passage will be observed \cite{Strogatz}. We estimate the duration of the $x$ wave as a function of $k_{a1}$ by calculating $y$ from zero to $+\infty
$, i.e.,
$$ T_{x}= \int_0^{\infty} \frac{dy}{(k_{a1}-k^c_{a1})x^{(3)}+c(y-y^{(1)})^2}= \frac {\pi}{\sqrt{c(k_{a1}-k^c_{a1})x^{(3)}}}.$$

Considering the inhibition term $-xy$ in Eq.~2a, if $y$ is near $y^{(2)}$, then $x$ can reach $x^{(3)'}=\frac{1+\sqrt{1-4j^2_1k^{'2}_1}}{2k_1}$, where $k^{'}_1=k_1+\frac{1-\sqrt{1-4j^2_2k^2_2}}{2k_2}$. When $k_{a1}\geq
k_{a1}^c=|H(y^*)|/x^{(3)'}$, $y$ can be triggered and activated, while $k_{a1}^c$ is mainly determined by parameters $j_1$, $k_1$, $j_2$, and $k_2$.

The envelope curve of Figure \ref{clustering01}B is based on this analysis, and is consistent with our simulations. Using the selected parameter set, we calculated the value of $k_{a1}^c$ based on the above estimation, and found $ k_{a1}^c =5.06\times 10^{-4}$, (i.e., very close to $0.001$), which shows that the ghost effect in the x wave exists.

In the early M phase, $y$ is activated and then evolves to $y^{(3)}$, and the repression term $-xy$ in Eq. \ref{xyequations}a makes $x^{(3)}$ unstable and represses $x$ to zero. A similar mechanism occurs in the late M phase, where the activation of $z$ is triggered by $k_{a2}y$, and the long duration of the $y$ wave is caused by the saddle-node remnant, i.e., the ghost effect. If we ignore the inhibition term $-zy$ from $z$ to $y$ in Eq. \ref{nondimensionalequ}b, then $k_{a2}^c = |I(z^*)|/y^{(3)}$ and $I(z^*)$ is the minimum of the $I(z)$ curve determined by $k_s$, $j_3$, and $k_3$, while $y^{(3)}$ is determined by $j_2$ and $k_2$. The duration of the $y$ wave is determined by $T_y\propto (k_{a2}-k^c_{a2})^{-1/2}$, and $k_{a2}^c$ is determined by $k_s$, $j_3$, $k_3$, $j_2$, and $k_2$. Again, using the selected parameter set, we calculated the value of $k_{a2}^c$ and found $ k_{a2}^c = 5.06\times 10^{-4}$, a value very close to
$0.001$.

\section*{Methods}

In this section, we develop methods to illustrate local stability along an evolving dynamic trajectory, then apply those methods to the yeast cell cycle process. Using the perfect parameter set, the perfect cell cycle trajectory evolving from the excited G1 state to the stable G1 state is shown in Figure \ref{wtcellcycle}A and B. The figure also shows $x(s)$, $y(s)$, and $z(s)$ in the upper panels of Figure \ref{jacob1}B, where $s$ is the trajectory length from the initial state. Note that $\Vd(t)$ and $\Vd(s)$ are the velocity of the trajectory as a function of time $t$ and of $s$, respectively. The normal plane of the trajectory is defined as the plane that is always perpendicular to $\Vd$.

\subsection*{Local spherical surface perturbation}

Points on a small spherical surface of radius $r_0$ around the exited G1 state were taken as perturbed initial states. In Figure \ref{jacob1} and Figure \ref{jacob2}, we set $r_0=0.01$. This sphere of perturbed states evolves together along the perfect trajectory with varying deviations, leading to a change in the shape of the sphere. At each time $t$ or trajectory length $s$, the deviation of each point on the perturbed sphere from the perfect state at the center was calculated as $r_i(t)$ or $r_i(s)$, where $i$ is the index of the perturbed trajectory. The normal radius $r_n$ of the perturbed sphere is the average of all $r_i$ for each value of $i$ on the cross section of the sphere that goes through the sphere center and is normal to $\Vd$. The forward and backward tangential radius $r_{t1}$ and $r_{t2}$ are defined as the farthest distance on the perturbed sphere from the center, in the direction parallel and antiparallel to the velocity $\Vd$, respectively. As an example, the normal radius $r_n(t)$ and $r_n(s)$ of the perfect yeast cell cycle trajectory are plotted in the middle panels in Figure \ref{jacob1}.

\subsection*{Local velocity analysis}

In the above method, the manifolds are formed by the trajectories evolving from the given small spherical surface at $t=0$ or $s=0$. In order to fairly examine the system's dynamic properties at different $t$ and $s$, circular perturbations on the normal plane of the yeast cell cycle trajectory were added at each step of the trajectory. The velocity vector of a perturbed trajectory can be decomposed with respect to the $\Vd$ axis into the normal component $V_{\bot}$ and the tangential/parallel component $V_{\|}$. The average velocities on the perturbed circle were then calculated. The magnitude of the average velocities $\overline{V}$ indicates how fast the manifold evolves, while the average of the normal components $\overline{V}_{\bot}$ and
tangential components $\overline{V}_{\|}$ denote how the manifolds converge and disperse on average. The $\overline{V}_{\bot}(t)$ and $\overline{V}_{\bot}(s)$ of the perfect cell cycle trajectory are plotted in the middle panels of Figure \ref{jacob1} as an example.

\subsection*{Eigenvalues of the normal Jacobian matrix}

The stability of any fixed point can be examined through the eigenvalues and eigenvectors of the Jacobian matrix of the ODEs. For a 3-dimensional dynamic system, $dx_i/dt =f_i(x_j)$, $i=1,2,3$, the Jacobian matrix is defined as
$$
 \left(
 \begin{array}{c c c}
   \frac{\partial{f_1}}{\partial{x_1}}& \frac{\partial{f_2}}{\partial{x_1}} & \frac{\partial{f_3}}{\partial{x_1}}\\
   \frac{\partial{f_1}}{\partial{x_2}}& \frac{\partial{f_2}}{\partial{x_2}} & \frac{\partial{f_3}}{\partial{x_2}}\\
   \frac{\partial{f_1}}{\partial{x_3}}& \frac{\partial{f_3}}{\partial{x_1}} & \frac{\partial{f_3}}{\partial{x_3}}
   \end{array}
   \right)
$$
If any real part of the eigenvalue of the Jacobian matrix is positive, the fixed point will be unstable in the direction of the correlative eigenvector. In other words, a fixed point remains stable only when all of its eigenvalues are negative.

Along the evolving trajectory, all state points should evolve in the velocity direction except for the final fixed point. To analyze the convergence properties of each manifold along the yeast cell cycle trajectory, the Jacobian matrix should be projected onto the normal plane of the trajectory. We adopt two unit orthogonal vectors
$\nd_1$ and $\nd_2$, together with unit velocity vector $\Vd/|\Vd|$, to compose a set of orthogonal unit vectors. Then the projecting matrix $P$ can be defined as $ (\nd_1 ,\nd_2, 0)$. Through the transaction of projection, the normal Jacobian matrix $J_{\bot}$ is equal to $P^T J P$, in the form
$$
 \left(
 \begin{array}{c c c}
   j_{11} &  j_{12} & 0\\
   j_{21} &  j_{22} & 0\\
   0 & 0 & 0
   \end{array}
   \right)
$$
The four non-zero elements consist of a $2\times 2$ normal Jacobian matrix whose eigenvalues $\lambda_1$ and $\lambda_2$ and eigenvector dictate the convergence properties in the normal plane of the evolving trajectory.

As an example, the real parts of $\lambda_1$ and $\lambda_2$ along the perfect cell cycle
trajectory are shown in the lower part of Figure \ref{jacob1}. The real parts are almost always negative except for the first part of the cell cycle and near the two vertex in Figure \ref{wtcellcycle}B. At the two vertex, where $x$ and $y$ reach their respective maximums and begin to decrease, there are positive peaks followed by large negative eigenvalues. Because the perfect cell cycle trajectory changes its evolving direction quickly near these vertices, we assume the positive eigenvalue peaks are caused by a numerical error in the velocity directions at the vertices.

\section*{Results}

\subsection*{The yeast cell cycle trajectory driven by saddle-node bifurcations with ghost effects is globally attractive}

Using the perfect parameter set, we depict the perfect yeast cell cycle trajectory in Figure \ref{wtcellcycle}A and B, and its local dynamic analysis results in Figure \ref{jacob1}. In Figure \ref{jacob1}, we plot the normal radius $r_n$ to describe the deviation from the perfect trajectory in the normal direction of the velocity of the initial state, the local average velocity $\overline{V}$ to depict the velocities of the circular perturbations along the evolving trajectory, and the real parts of eigenvalues $\lambda_1$ and $\lambda_2$ to describe the convergence properties in the normal plane of the trajectory. These results reveal some interesting dynamic properties along the perfect cell cycle trajectory.

In the first part of the trajectory (G1 and S phases), Figure \ref{jacob1} shows that $r_n$ reaches a peak at $t=4.82$ and $s=5.32$, suggesting that the perfect trajectory diverges at the beginning of the S phase. When the system evolves at $t=20.09$ and $s=8.59$ to the first vertex ($x_{max}=4.33$, $y=0.03$, $z=0.00$) in Figure \ref{wtcellcycle}B, $x$ is fully activated and has repressed $z$ to zero, and $x$ also begins to trigger the activation of $y$ by saddle-node bifurcation. The duration of the $x$ wave is given by $T_x\propto (k_{a1}-k^c_{a1})^{-1/2}$. Because $ (k_{a1}-k^c_{a1})\ll 1$ in the perfect parameter set, there is a ghost effect with long duration of $x$ wave. The results in Figure \ref{jacob1} show that, near the first vertex, the normal radius $r_n$ decreases to near zero, the real parts of eigenvalues $\lambda_1$ and $\lambda_2$ sharply decrease to negative values, and the average local velocity $\overline{V}\approx 0$. These results demonstrate that the local manifold of the cell cycle trajectory converges to the first vertex, and that the trajectory evolves very slowly near the first vertex  because of the ghost effect. Note that the first vertex state corresponds to the DNA replication event. This suggests that the ghost effect after bifurcation provides a sufficiently long duration for the $x$ wave to execute a DNA replication event, while the converging manifold and slowly evolving trajectory near the first vertex provides a suitable state for the DNA replication checkpoint. On the other hand, the vertex with a convergent manifold works as a hidden intrinsic checkpoint mechanism for the cell cycle process.

A similar dynamic property is found in the later part of the perfect cell cycle trajectory. When $t=39.16$ and $s=10.78$ in the early M phase, the trajectory diverges. Near the second vertex (0, $y_{max}=4.35$, 0), $y$ is fully activated and has repressed $x$ to zero, so $y$ is triggering the activation of $z$ through saddle-node bifurcation. This saddle-node bifurcation and ghost effect lead to a converging, attractive and slowly evolving manifold near the second vertex. The second vertex corresponds to the spindle assembly and separation event.

\subsection*{Modularity of state and parameter space in the perfect cell cycle model}

The modularity of both state and parameter space can be found in the perfect cell cycle model. In Figure \ref{wtcellcycle}A and B, not only do the $x$, $y$, and $z$ waves have little overlap time, the three vertices in the phase map have sharp pointed ends and the kinetic parameters in the $x$, $y$, and $z$ waves are also decoupled into independent groups.

Near the first vertex, $k_{a1}^c=|H(y^*)|/x^{(3)}$, where $H(y^*)$ is the minimum of $H(y,j_2,k_2)$ and $x^{(3)}=\frac{1+\sqrt{1-4j^2_1k^{2}_1}}{2k_1}$. So the first bifurcation near the first vertex ($x_{max}$, 0, 0) is mainly controlled by the $y$ wave's kinetic parameters $j_2$ and $k_2$ through $H(y^*)$, and the $x$ wave's kinetic parameters $j_1$ and $k_1$ through $x^{(3)}$. Note that these values are almost independent of the $z$ wave's kinetic parameters because the event occurs mainly in the $x-y$ plane with $z=0$. Similarly near the second vertex, $k_{a2}^c=|I(z^*)|/y^{(3)}$, where $I(z^*)$ is the minimum of $I(z,j_3,k_3,k_s)$ and
$y^{(3)}=\frac{1+\sqrt{1-4j^2_2k^{2}_2}}{2k_2}$. The second bifurcation near the second vertex (0, $y_{max}$, 0) is mainly in the $y-z$ plane with $x=0$, and is controlled by the $z$ wave's kinetic parameters $j_3$ and $k_3$, and by $k_s$ through $I(z^*)$ and the $y$ wave's kinetic parameters $j_2$ and $k_2$ through $y^{(3)}$. Again, these values are almost independent of the $x$ wave's kinetic parameters.

When $0<(k_{a1}-k_{a1}^c)\ll1$ and $0<(k_{a2}-k_{a2}^c)\ll1$, the relative saddle-node bifurcations occur in sufficient order. The duration of $x$ wave is mainly controlled by the $(k_{a1}-k^c_{a1})$ and the maximum of $x$ wave $x^{(3)}$ shown in \ref{clustering01}B. Similar is the $y$ wave, in which the duration of $y$ wave is mainly controlled by the $(k_{a2}-k^c_{a2})$ and the maximum of $y$ wave $y^{(3)}$. In this case, the system not only provides a long duration for the execution of the DNA replication and mitosis events, but also separates the $x$, $y$, and $z$ waves into relatively independent events and decouples the kinetic parameters of different waves and phases. These results provide a suitable strategy for the regulation of sequential events, especially because the change and modification parameters of one event influence other events very little.

\subsection*{The perfect cell cycle model and its trajectory}

The robustness and manifold analysis along the cell cycle trajectory provide more dynamic information of the cell cycle process. Novak and coworkers discussed a kinetic differential equation model in which the budding yeast cell cycle trajectory could be separated into excitation and relaxation periods according to the eigenvalues of the Jacobian of the trajectory \cite{Novak2006}. The divergence and convergence manifold has also been observed in a Boolean network model, where the manifold converges at the S phase state corresponding to a DNA replication checkpoint \cite{Li2004}. In the perfect cell cycle model we find some new results, including the manifold of key regulators diverges and converges in different cell cycle phases, and the dynamic robustness and modularity of cell cycle process.

Let us interpret the results from the biological perspective. Suppose that the states (activities of the key regulators) of yeast cells can be dynamically observed, if a group of yeast cells start from different excited G1 states with fluctuations of biochemical parameters that tend to vary from cell to cell, how do the yeast cells evolve during the whole cell cycle process? During the early S phase, the states of the yeast cells should be observed to separate and diverge significantly (diverge manifold) and their states are changing quickly (large average velocity), and in this phase the yeast cells activate kinases and express genes for the execution of DNA synthesis and replication task. When the cells enter the late S phase that the DNA replication task is working, the yeast cell states begin to change slowly and gradually converge to the same point; this corresponds to the system evolves into the first vertex $(x_{max},0,0)$, there is long duration with decreased velocity and convergent manifold. Similarly, in the early M phase the yeast cell states separate, diverge and change quickly. When the metaphase/anaphase transition with spindle assembly and separation task takes place, where the system near the the second vertex $(0, y_{max},0)$, the yeast cell states should change slowly and converge again. Finally, all the yeast cells should evolve to the stable G1 state $(0, 0, z_{max})$ waiting for another cell cycle signal. During the whole cell cycle process, from the DNA replication event to the mitosis, the dynamic robustness and modularity of the system provide a suitable strategy to execute the sequential events reliably. So in cell cycle process, the manifolds diverge and converge, wave after wave, cycle in cycle out. This is the dynamical picture of cell cycle.

The durations of the $x$ and $y$ waves in the perfect cell cycle model are controlled by, and sensitive to, $k_{a1}$ and $k_{a2}$, respectively. In the SI we investigate a simplified cell cycle model with inhibitor $I$, in which $I$ represses $y$ and $x$ represses $I$ by one-step or multi-step phosphorylation, noted as $x \dashv I\dashv y$. In the SI, we show that the inhibitor with phosphorylation can largely reduce the duration sensitivity of the $x$ wave $T_x$ to changes in the relative kinetic parameters from the $x$ to inhibitor $I$, and the $x$ wave duration $T_x$ is not sensitive to the maximum of $x$ wave $x^{(3)}$. These differences can be also used to distinguish the direct activation and indirect activation (with inhibitor) between the successive waves.

\subsection*{The ``ideal" yeast cell cycle trajectory and the cell cycle checkpoints}
In a real yeast cell cycle process, yeast cells produce new daughter cells through DNA replication in the S phase and undergo mitosis in the M phase, with each event associated with a relative checkpoint mechanism. For example, when the cell is in a DNA replication event, or the DNA is damaged in the S phase, the relative pathway is activated and turns on the DNA replication checkpoint to keep the cell in the S phase; only when the damaged DNA is repaired or the yeast cell finishes the DNA replication event is the DNA replication checkpoint turned off and the the yeast cell allowed to enter the G2 and M phases.

To simulate the above cell cycle checkpoint function, we artificially changed  $k_{a1}$ and $k_{a2}$ to represent the on or off states of the cell cycle checkpoints. When the DNA replication checkpoint is turned on, $k_{a1}=0$; when it is turned off, $k_{a1}>k_{a1}^c $. Similar to the spindle checkpoint, $k_{a2}=0$ represents the on state while $k_{a2}>k_{a2}^c $ is the off state. We denote this as the real or ``ideal" yeast cell cycle trajectory. The ideal trajectory evolves through the 3 vertex states $(x^{(3)}, 0, 0)$, $( 0, y^{(3)}, 0)$, and $( 0,  0,  z^{(3)})$ in order, and the checkpoints not only control the duration of the $x$ and $y$ waves, but also separate the $x$, $y$, and $z$ waves. Furthermore, the checkpoint mechanisms make the kinetic parameters in different waves independent. This kind of extrinsic checkpoint mechanism in real yeast cell cycle process provides a more stable duration for each phase.

The ideal yeast cell cycle trajectory is globally attractive and stable. The following discussion provides further evidence for  this conclusion. If we were to ignore the activations and repressions among $x$, $y$, and $z$ in Eq.~\ref{nondimensionalequ} as formulated in the Model section of this work, there would be 3 fixed points, one on each of the $x$, $y$, and $z$ axis. Thus, $x^{(1)}=0$, $x^{(3)}$, $y^{(1)}=0, y^{(3)}$, and $z^{(1)}=0, z^{(3)} $ are stable nodes, while $x^{(2)}$, $y^{(2)}$, and $z^{(2)}$ are unstable saddles. Consider the repression terms $-xy$, $-yz$, and $-k_ixz$ in Eq. \ref{nondimensionalequ}, where only one of $x$, $y$, and $z$ can be in its maximum state, and $(0,0,0)$ is unstable. In this case, only $(x^{(3)}, 0, 0)$, $(0, y^{(3)}, 0)$, and $(0, 0, z^{(3)})$ are the possible final stable attractors.

When discussing the evolution of the cell cycle engine, Andrew Murray suggested to make ``toy systems" that are much simpler and mimic the key features of cell cycle process for the deeper understanding of cell cycle\cite{Murray2004}. From the intrinsic checkpoint mechanism in the perfect cell cycle to the outside checkpoint mechanism in real yeast cell cycle process, our results suggest a possible evolution course of cell cycle network and sequential-task regulatory networks.

\subsection*{Imperfect yeast cell cycle trajectory}

\begin{figure}[!ht]
\begin{center}
\centerline{\includegraphics[width=.7\textwidth]{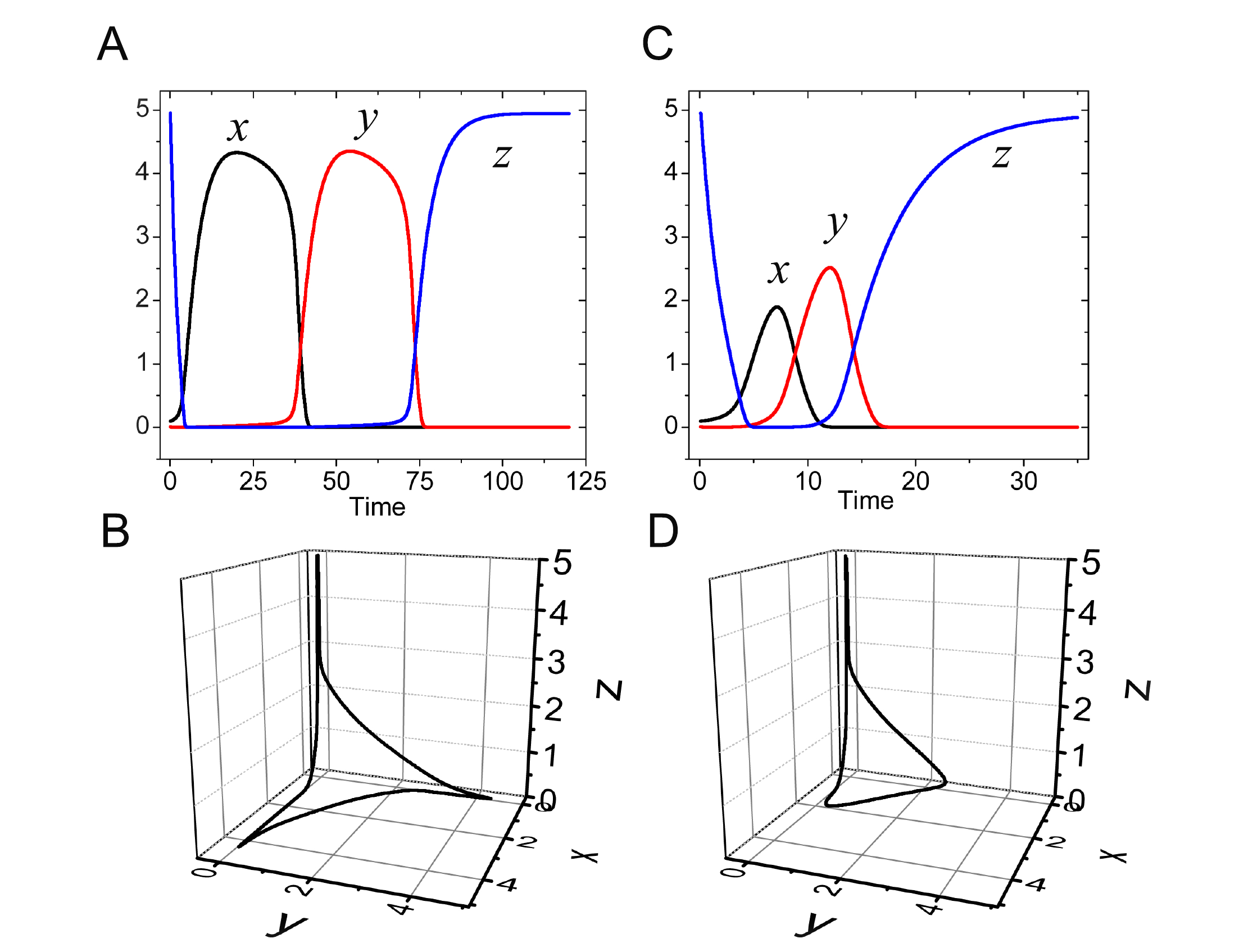}}
\caption{\label{wtcellcycle} The evolving cell cycle trajectory. (A) and (B) Using the perfect parameters ($k_{ai}=0.001$). (C) and (D) Using the imperfect parameters ($k_{ai}=0.04$). In panel B, the first vertex ($x_{max}$, 0, 0) corresponds to the DNA replication event, while the second vertex (0, $y_{max}$, 0) corresponds to the spindle assembly and separation event. }
\end{center}
\end{figure}

\begin{figure}[!ht]
\begin{center}
\centerline{\includegraphics[width=0.8\textwidth]{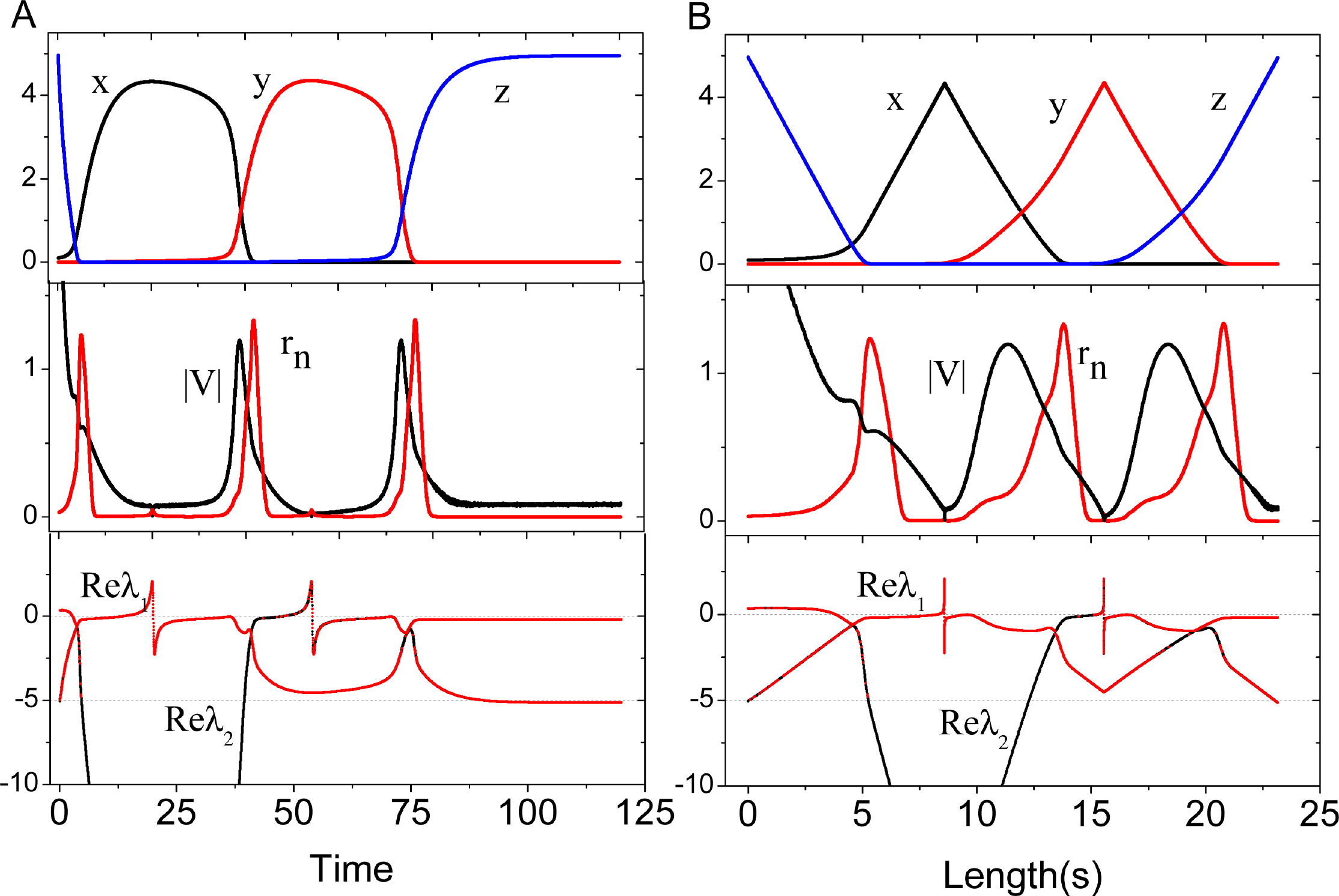}}
\caption{\label{jacob1}Local dynamic analysis of a perfect cell cycle trajectory ($k_{ai} = 0.001$) as (A) a function of time and (B) a function of curve length $s$.  The upper panels plot the evolving trajectory. The middle panels plot the normal radius $r_n$ of the local spherical perturbation surface (red curve) and the average local velocity $|\overline{V}|$ (black curve), showing that the manifolds converge to a small state space when $x$ or $y$ are fully activated. The lower panels plot the real parts of eigenvalues $\lambda_1$ and
$\lambda_2$ of the normal Jacobian matrix. }
\end{center}
\end{figure}

\begin{figure}[!ht]
\begin{center}
\centerline{\includegraphics[width=0.7\textwidth]{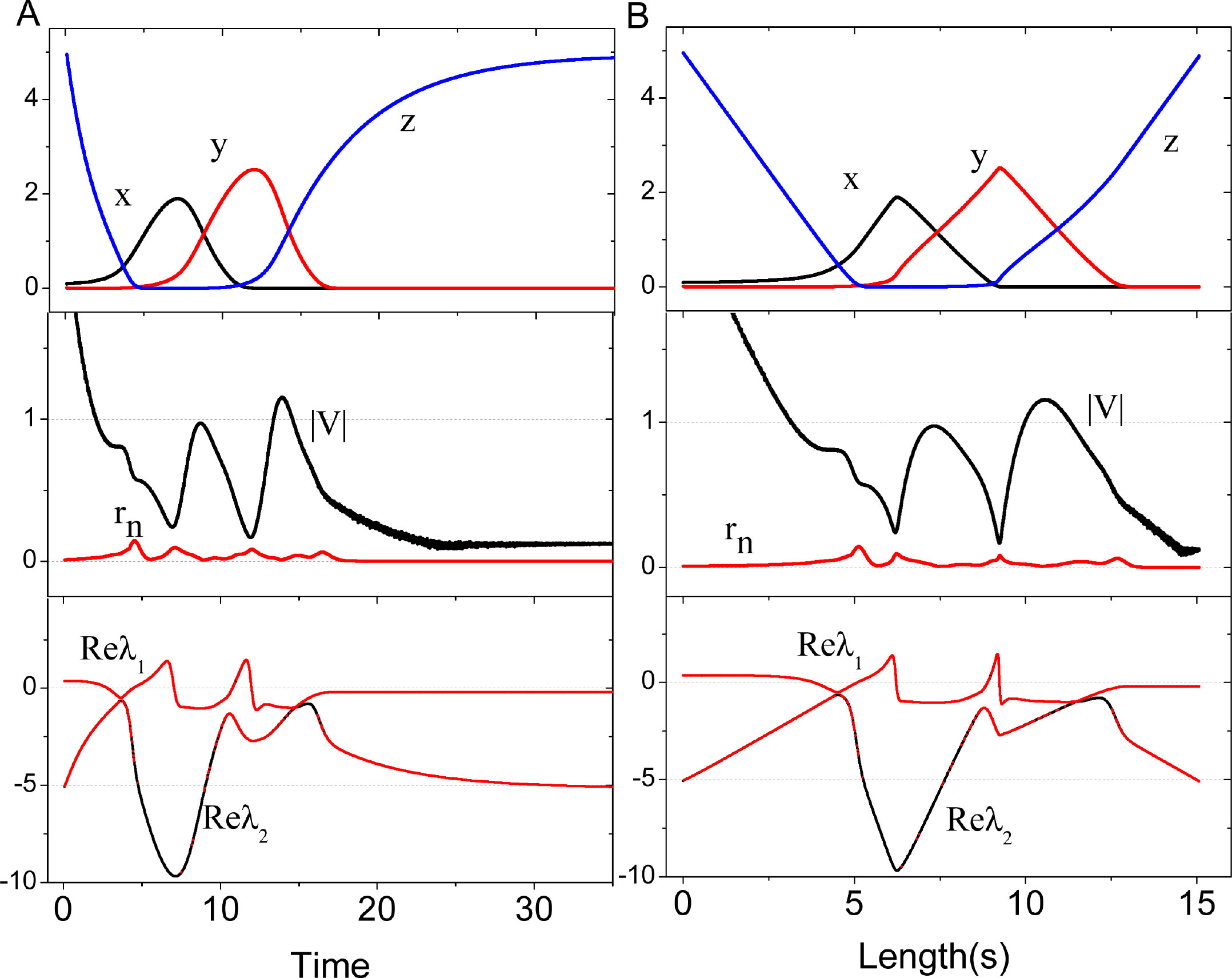}}
\caption{\label{jacob2}Local dynamic analysis of an imperfect cell cycle trajectory ($k_{ai}=0.04$). The upper panels plot the evolution trajectory. The middle panels plot the local spherical surface perturbation normal radius $r_n$ (red line) and the average local velocity $|\overline{V}|$. The lower panels plot the real parts of the eigenvalues $\lambda_1$ and $\lambda_2$ of the normal Jacobian matrix.}
\end{center}
\end{figure}

In this section we investigate a counter example, the imperfect yeast cell cycle process, where we set $k_{a1}=0.04$ and $k_{a2}=0.04$, and $k_{a1}>k_{a1}^c$ and $k_{a2}>k_{a2}^c$, all of which are relatively far from the bifurcation points. The imperfect trajectory (Figure~\ref{wtcellcycle}C and D) and its local dynamic properties (Figure~\ref{jacob2}) are quite different from the above perfect yeast cell cycle trajectories. In Figure~\ref{wtcellcycle}C and Figure~\ref{wtcellcycle}D, the maxima of $x$ and $y$ just reach half of $x^{(3)}$ and $y^{(3)}$, there is a relatively large overlap time for the $x$, $y$, and $z$ waves, and there is no sharp pointed end near the 1st and 2nd vertices in phase space. The duration times of the $x$ and $y$ waves are only 1/6 that of the duration times of the perfect cell cycle trajectory. All of these results show that the $x$, $y$, and $z$ waves are coupled together. In Figure \ref{jacob2}, one observes that the expansion is less than the perfect trajectory ($r_n$), even though the local manifold also expands and converges along the trajectory. Near the vertices in the phase map, the manifold cannot converge to a narrow state space with a certain evolving velocity ($\overline{V}$), which then cannot provide a long enough duration for DNA replication and spindle assembly and separation events. Furthermore, there is no suitable point for the checkpoint mechanism.

\section*{Discussion and Conclusion}


The yeast cell cycle process executes DNA replication and mitosis in subsequent order, and should be robust against any intrinsic fluctuations and environmental change. In this paper, we constructed and investigated a simplified yeast cell cycle model as a toy model to capture the dynamic essence of sequential-task biological processes. We have shown that the cell cycle process is an excitable system with global robustness in the state space, and that the phase transitions are driven by dynamic bifurcations containing ghost effects. These bifurcations and associated ghost effect sufficiently provide the longer durations required for each event or wave, as well as the modularity of the state and parameter spaces. This cell cycle model is robust and modular in both state and parameter space. The robustness in state space ensures that the cell cycle process is stable against fluctuations in protein activity. The modularity and decoupling of the kinetic parameters in the different waves avoids allowing small parameter changes in any one event to influence the general cell cycle process. Our wave transition model provides an efficient and easily controlled strategy for executing the cell cycle processes.

Furthermore, our toy cell cycle model suggests a possible synthetic network design for robustly executing other sequential-task processes. In the synthetic network, each event can be controlled by the relevant key regulators, and the duration of each event is regulated by the activation strength between successive events (similar to $k_{a1}x$ and $k_{a2}y$ in our toy model). These synthetic networks could be used to check our prediction that the long duration and modularity of the sequential-task process is caused and controlled by the ghost effect.

Most of the insights we obtained from the toy model are independent of the specific formulation of the model and the number of dimensions, and the sufficient condition is that the dynamic cell cycle models should be governed by sequential saddle-node bifurcations containing ghost effects. We believe that our results and analysis method are applicable to other eukaryotic cell cycle processes and other sequential-task biological processes. Life is modular in parts and more than the sum of its parts, and this robustness and modularity are the design principles for biological regulatory networks \cite{Hartwell-Module-1999, Kitano2002, Sneppen2005, Alon2007}. Indeed, if a network behaves along a globally attractive trajectory as well as being modular in state and parameter space, it can execute successive irreversible events more robustly and effectively and thus obtain an evolutionary advantage.



\section*{Supporting Information}
{\bf{Text S1}} This file contains details needed to understand the main body of this work, which are arranged as follows:

I The regulatory network in budding yeast cell-cycle process and a 3-node model

II A simplified cell-cycle model with inhibitor

{\bf {Figure S1}}  The regulatory network of key regulators in the budding yeast cell-cycle process, which can be separated into the G1/S, early M, and late M modules. It is contained within Section I of the SI.

{\bf {Figure S2}}  The network of the one-step phosphorylation inhibitor that is inserted between $x$ and $y$ in the 3-node yeast cell-cycle model. It is contained within Section II of the SI.

{\bf {Figure S3}} An inhibitor with 4-step phosphorylation is added between $x$ and $y$ in the yeast cell-cycle model, which decreases the sensitivity of wave duration to changes in the triggering rate $b$. It is contained within Section II of the SI.

\section*{Acknowledgments}
The authors are grateful to Chao Tang, Louis Tao, Tiejun Li, Xiaomeng Zhang, Tianqi Liu, and Yuhang Hou for helpful discussions. The work is supported by NSFC grants nos. 11174011, 11021463, and 91130005 (F.Li), and nos. 11074009 and 10721463 (Q.Ouyang).

\section*{Figure Legends}

\end{document}